\documentclass[a4paper, amsfonts, amssymb, amsmath, reprint, showkeys, nofootinbib, twoside]{revtex4-1}
\usepackage[english]{babel}
\usepackage[utf8]{inputenc}
\usepackage[colorinlistoftodos, color=green!40, prependcaption]{todonotes}
\usepackage{amsthm}
\usepackage{mathtools}
\usepackage{physics}
\usepackage{xcolor}
\usepackage{graphicx}
\usepackage[left=23mm,right=13mm,top=35mm,columnsep=15pt]{geometry} 
\usepackage{adjustbox}
\usepackage{placeins}
\usepackage[T1]{fontenc}
\usepackage{lipsum}
\usepackage{csquotes}
\usepackage{bm}
\usepackage[pdftex, pdftitle={Article}, pdfauthor={Author},colorlinks=true, urlcolor=blue, pdffitwindow=true, linkcolor=blue, citecolor=blue, pdfstartview={FitH}]{hyperref} 
\usepackage{epstopdf}
\usepackage{graphicx}
\begin{document}
\title{Spinor quantum states of the Dirac's core/shell at fm-space}

\author{Sami Ortakaya}
    \email[Correspondence email address: ]{sami.ortakaya@yahoo.com}
    \affiliation{Ercis Central Post Office - 65400 Van, Turkey }
     \affiliation{Institute for Globally Distributed Open Research and Education (IGDORE), Gothenburg, Sweden}

\date{\color{blue}\today} 

\begin{abstract}
In this work, we present a model for the behavior of Dirac particles under the tensor effect in the spherical core/shell regime. We examine the change of energy levels corresponding to the particles localized in a space of approximately 1. 0 fm in the core region of the quantum sphere, with the well width. It also occurs from the analytical solutions that the two different levels accompany particle states of the same mass. Additionally, the solutions exhibiting anomalous behavior, giving rise to antiparticle-type states, occur at heavier mass.
\end{abstract}


\maketitle

\section{Introduction} \label{sec:outline}
As a cornerstone of solvable models in relativistic quantum mechanics, the Dirac equation boasts a rich legacy of applications, as evidenced by its extensive study in high-energy physics \cite{r12, r13, r14}, optical topics \cite{o1} and condensed matter physics \cite{c1}. Theoretical approaches to modelling Dirac's particles have, for 50 years, focused on the interactions in view of the spinor systems \cite{r1, r2, r3, r4}. Such models have been launched on the formation of quantum systems such as hot nucleus \cite{a1}, correlation in nuclear spinors \cite{a2} and polar representation \cite{a3}. Within the context of space-time dimensionality, exact solutions of the Dirac equation have been also obtained \cite{rev1, rev2, rev3}.

In the computational manner, pseudo-spin symmetric solutions of Dirac equation based on radial interactions in spherical shell have pioneering results in mathematical physics \cite{aydogdu}. In studies involving the integrated effect of tensor interactions regarding spherical quantum wells within the Dirac equation, degenerate states and their removing have been observed in the analytical studies such as exponential oscillator \cite{ref1}, Yukawa tensor interaction \cite{ref2} and  Coulombic tensor \cite{ortakaya2013}. Furthermore, Dirac particles in quantum well with topological insulator \cite{qwell1} and spherical core systems \cite{qwell2} have been also studied through analytical approaches. In previous research we have demonstrated that the pseudo-spin solutions in Dirac's spinor systems are calculable context on the relevant energy spectra through spatially varying mass \cite{sami2013a}.

In the Dirac equation, the pseudospin concept occurs in the constant potential energy context \cite{pss}. In total, we deal with the radial interactions are given by 
\begin{equation}
\Sigma(r)=V(r)+S(r);\quad \Sigma'=0,
\end{equation}
where $V(r)$ and $S(r)$ are the scalar potential and the relevant mass distribution, respectively. Constant potential energy allows the radial-spinor equations to be reduced to Schrödinger-type solvable eigenvalue equation. In a way, under the spin \& pseudospin concepts, the upper- and lower-spinor states are clearly revealed when the analytical or full numerical method is applied.

In radial form, the Schrödinger-type solvable spinor equations require a constant total-potential under the pseudospin concept, so the other "difference potential, $\Delta (r)=V(r)-S(r)$" is valid for variable form through radial mass distribution. This "key mechanism" also needs to be understood in the spherical quantum well, where the size effect is described. In other words, in cases where the mass distribution varies only in the core and shell regions, there is a need to determine how the energy values change with the size effect, similar to the formation of particles and antiparticles, even if it is constant in the material region, to determine normal or anomalous states.

In this study, we focus on calculation of energy levels regarding Dirac particles in the core/shell sphere, considering quantum well regions with different mass distributions. As a logical approach to the quantum confinement, we determine the particle and antiparticle states representing the decrease and increase in energy levels based on the increasing change in the core radius. From these results, we also establish that the heavier mass distribution in the core region corresponds to an antiparticle state. We also show new numerical results of the tensor interaction related to physically acceptable solutions applied to the spherical quantum well at fm-scale nuclear distance.
\section{Modeling} \label{sec:develop}
    Considering atomic units $\hbar=c=1$, a typical Dirac equation with spatial varying mass including tensor, is given by \cite{sami2013a}
    \begin{eqnarray}
        \left[\bm{\alpha}\cdot\bm{p}+\beta m(r)-{\rm i}\beta\bm{\alpha}\hat{r}U(r)\right]\Psi(\bm r)=[E-V(r)]\Psi(\bm r),\nonumber\\
        &&
    \end{eqnarray}
    where $\bm{p}={\rm i}\hbar\bm{\nabla}$ is the momentum operator, $m(r)$, $U(r)$ and $V(r)$ denotes position-dependent mass which has energy equivalent, tensor interaction and spherical symmetric potential, respectively. $\bm{\alpha}$ and $\beta$ are also Dirac matrices defined by
 \begin{figure}[!hbt]
\centering
\scalebox{.5}{\includegraphics{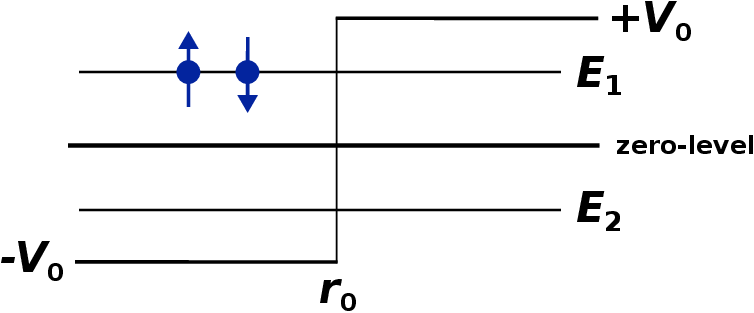}}
\caption{Scalar potential energy, $V(r)$ profile of Dirac's core/shell}\label{fig:1}
\end{figure}
\begin{equation} \bm{\alpha}=
  \left( {\begin{array}{cc}
   0 & \bm{\sigma} \\
   \bm{\sigma} & 0 \\
  \end{array} } \right), \qquad   \bm{\beta}=\left( {\begin{array}{cc}
   0 & I \\
   I & 0 \\
  \end{array} } \right),
  \end{equation}
  where $I$ is $2\times 2$  identity matrix and $\bm{\sigma}$ represents three-vector spin matrices via following spinor form:
    \begin{equation} \Psi_{n\kappa}(\bm{r})=
  \left( {\begin{array}{c}
   \frac{F_{n\kappa}(r)}{r}\Omega_{jm}^{\ell}(\theta,\,\varphi) \\
   \frac{G_{n\kappa}(r)}{r}\Omega_{jm}^{\tilde{\ell}}(\theta,\,\varphi) \\
  \end{array} } \right),
  \end{equation}
where $F$ and $G$ are upper- and lower- spinors; respectively, $\Omega$ is the spin \& pseudospin spherical harmonics for $\ell$ and $\tilde{\ell}$. Within the spherical nuclei, the eigenvalues of spin-orbit coupling lies that
 \[
    \kappa = \bigg\{\begin{array}{lr}
        -\left(j+1/2\right)<0,&  j=\ell+1/2\quad \text{(aligned spin)},\\
        \left(j+1/2\right)>0,
         & j=\ell-1/2 \quad\text{(unaligned spin)}.
        \end{array}
  \]
 We can launch the pseudospin symmetry as a case of potential energy profile, $m(r)=m_{i}+S(r)$ for $i=1$ in core and $i=2$ in shell; so considering that
\begin{subequations}
 \begin{eqnarray}
 &&\Sigma(r)=V(r)+S(r),\\
&& \Delta(r)=V(r)-S(r),
 \end{eqnarray}
we should get the spinor $G(r)$ in pseudospin represenatation at $\text{d}\Sigma(r)/\text{d}r=0$. One can also obtain two couples for upper- and lower-spinor components 
\begin{eqnarray}\label{sp}
&&\left(\frac{{\rm d}}{{\rm d}r}+\frac{\kappa}{r}-U(r)\right)F_{n\kappa}(r)=(m_{0}+E_{n\kappa}-\Delta)G_{n\kappa}(r),\nonumber\\
&&\left(\frac{{\rm d}}{{\rm d}r}-\frac{\kappa}{r}+U(r)\right)G_{n\kappa}(r)=(m_{0}-E_{n\kappa}+\Sigma)F_{n\kappa}(r),\nonumber\\
{}
\end{eqnarray}
so we obtain that the solvable Schrödinger-type equation in the pseudospin symmetry where radial varying energy becomes $\Sigma'=0$ in Equation (\ref{sp}). Inserting pseudospin symmetry, we should have a solvable eigenvalue equation of the form  
 \end{subequations}  
\begin{equation}\label{eq5}
\left[\frac{\text{d}^2}{\text{d}r^2}-\frac{\kappa(\kappa-1)+f[U(r)]}{r^2}-\epsilon_{i}\right]G(r)=0
\end{equation}
where
\begin{subequations}
\begin{eqnarray}
\epsilon_1=m_1^2-(E+V_0)^2,\quad r<r_0,\\
\epsilon_2=m_2^2-(E-V_0)^2,\quad r\geq r_0.
\end{eqnarray}
Putting Coulombic interaction as a radial component, $U(r)=-\frac{U_{0}}{r}$, we obtain that
\begin{equation}
f(U_{0})=(2\kappa-1)U_{0}+U_{0}^2.
\end{equation}
\end{subequations}
Defining the physical acceptable solution $G=r^{a}g(r)$ for $a>0$, we have
\begin{equation}\label{alpha}
a_{\pm}(\kappa)=0.5\pm\sqrt{0.25+\kappa(\kappa-1)+(2\kappa-1)U_{0}+U_{0}^2}.
\end{equation}
The second proposed-function based on the behavior of wave function at large distance reads \[g(r)=\exp(-b r)M(r),\] and then Equation (\ref{eq5}) is turn into the Kummer's eigenvalue equation is obtained the form
\begin{equation}
rM''+(2a-2b r)M'-2ab M=0,\quad b=\sqrt{\epsilon_i}>0.
\end{equation}
So that, unnormalized form is given in following function
\begin{equation}G(r)=r^{a}e^{-b r}M(a,\,2a\,;2b r),\end{equation}
where $M$ denotes confluıent hypergeometric functions.

\section{Numerical Results}

From Eq. (\ref{alpha}), we conclude that the degeneracies are taken throught $U_{0}=0$, so there is a degeneracy between $\kappa=0$ and $\kappa=-1$. The all degeneracies remove when tensor interaction exists and then we can take the values of $E_{n,\kappa}$ which has non-degeneracy as $E_{n,0}\neq E_{n,-1}$.

Especially, the mathematical manner at ground state; $\kappa=0$ leads to $$a(U_{0}=0)=a(U_{0}=1).$$ But now we consider the arbitrary values of $U_{0}$ and then we should know the range, $a>0$ under physical acceptable solutions. In the ground state $\kappa=0$, the $U_{0}$ has to been tuned in the range, $U_{0}^2<U_{0}$ for $a>0$ in Eq. (\ref{alpha}) through acceptable lines. In a way, we have the second solution given by
   \begin{equation}\label{alpha2}
a_{-}=0.5-\sqrt{0.25+U_{0}^2-U_{0}}\;\text{for}\; a>0\;\wedge\; U_0^2-U_0<0.
\end{equation}

We firstly set up energy spectra without tensor interaction, so the solutions of Eq. (\ref{eq5}) are obtained by the boundary conditions
\begin{eqnarray}
G_{1}(r_{0})&=&G_{2}(r_{0}),\nonumber\\
\frac{1}{m_{1}}G_{1}'(r_0)&=&\frac{1}{m_{2}}G_{2}'(r_0),
\end{eqnarray}
and then the considered transcendental equation yields energy spectra.

\begin{figure}[hbt!]
        \centering
\scalebox{.85}{\includegraphics{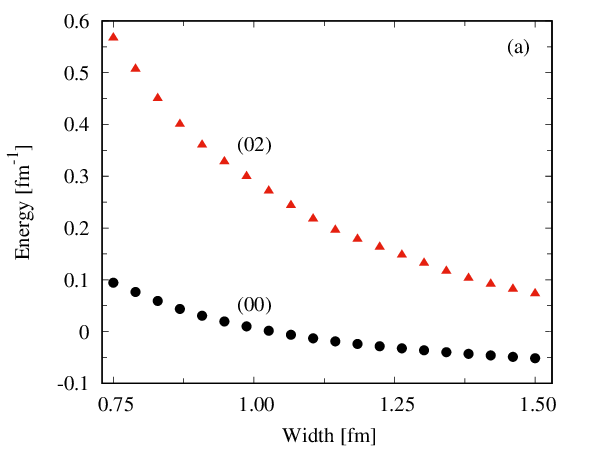}}\\  
\scalebox{.85}{\includegraphics{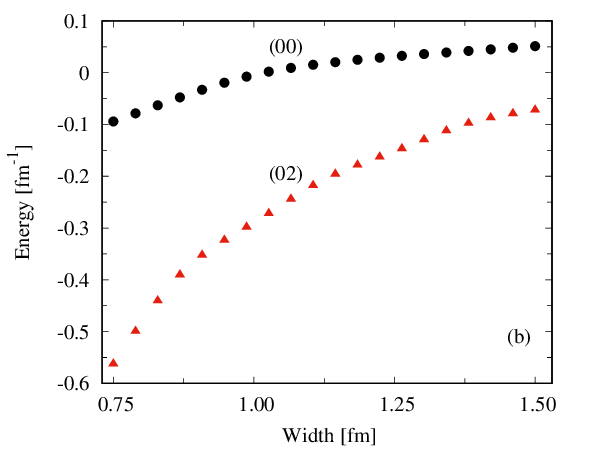}}\caption{Energy eigenvalues for (a) normal level (particle) and (b) anomalous level (antiparticle) related to $(n\kappa)$ states as a function of well width $r_0$ without tensor interaction.}
    \label{fig:enter-label1}
    \end{figure}

The energy eigenvalues as a function of the well width are shown in Figure \ref{fig:enter-label1}. The depth of the quantum well corresponds to $V_0 =1.0 \, {\rm fm^{-1}}$, taking $2\, {\rm fm^{-1}}$ in total and there is no tensor interaction. The increase in energy in the excited state and the decrease in energy with increasing well width represent particle states in view of light rest-mass $m_1 = 1.5\,{\rm fm^{-1}}$ and heavier value of $m_2 = 1.75\,{\rm fm^{-1}}$. The other behaviour represents antiparticles at heavier rest-mass $m_1 = 1.75\,{\rm fm^{-1}}$, so the shell layer is analyzed at a lighter effective mass $m_2 = 1.5\,{\rm fm^{-1}}$. For the particle case, the core region can be considered at a heavier effective mass. The energy spectra shown summarise the relationships between the particle and antiparticle states and the mass combination in a spherical core/shell structure.

\begin{figure}[hbt!]
        \centering
\scalebox{.75}{\includegraphics{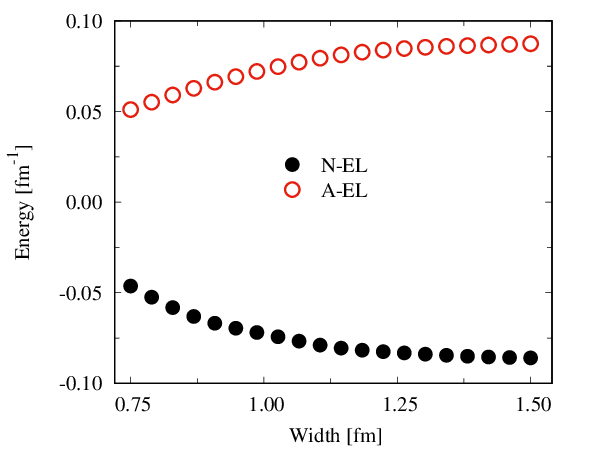}}\caption{Ground state energy as a function of well width $r_0$ in the presence of Coulombic tensor interaction $U_{0}=0.5$}
    \label{fig:enter-label2}
    \end{figure}

When the tensor parameter $U_0$ is unitless, it is obtained that $a_{\pm}=0.5$ for $U_0=0.5$. Here, the range $a>0$  provides physical acceptable condition given by 
\begin{equation}G(r)\propto r^a,\quad G(r=0)=0.\end{equation} 
The ground state ($n=0$, $\kappa=0$) energy eigenvalues for $U_0=0.5$ are shown in Figure \ref{fig:enter-label2} through heavy $(m_1 = 1.75\,{\rm fm^{-1}})$ and light $(m_1=  1.5\,{\rm fm^{-1}})$ rest-mass energies. A charged particle of mass $m_1 = 1.75\,{\rm fm^{-1}}$ is heavier mass in the core region and $m_2=  1.5\,{\rm fm^{-1}}$ denotes lighter mass in the shell layer, which can also be considered as "effective mass". On the other hand, the decreasing behavior of the energy spectra in the core/shell with increasing quantum-well width for $m_1=  1.5\,{\rm fm^{-1}}$  and $m_2=  1.75\,{\rm fm^{-1}}$ is in accordance with the normal energy level (N-EL) concept, similar to particle-state assignment. The heavier effective mass in the core layer ($m_1=  1.75\,{\rm fm^{-1}}$ and $m_2=  1.5\,{\rm fm^{-1}}$) has anomalous energy level (A-EL) in accordance with the antiparticle-state assignment, so the energy values increase with increasing well width. In particular, these results partially differ from those obtained for $U_0=0$ shown in Figure \ref{fig:enter-label1}. In the presence of the tensor interaction, the behaviour of the heavier-mass A-EL is from positive energies to higher energies. In the absence of the tensor potential, the A-EL (or antiparticle) shifts from negative to zero level, so that the probability of tunneling for the so-called A-EL antiparticle contexts through tensor interaction increases. 
    
    \section{Discussion and Conclusion}
    
\begin{figure}[hbt!]
        \centering
\scalebox{.75}{\includegraphics{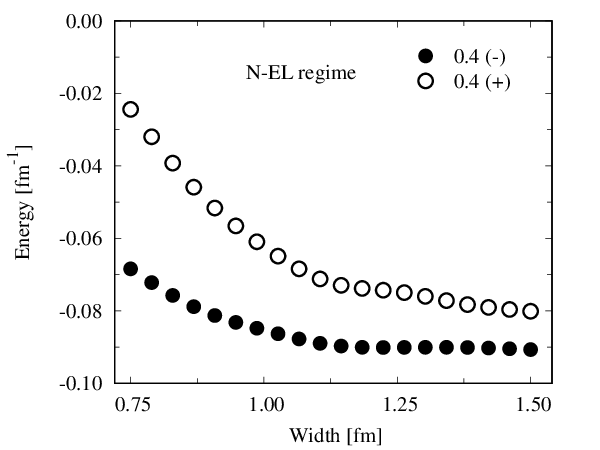}}\caption{Particle states and its behaviour as a function of well width at ground state for tensor interaction $U_0=0.4.$}
    \label{fig:enter-label3}
    \end{figure}

The key feature of analytical solutions is that the physically acceptable way is realized if $a_{\pm}>0$. For $U_0 = 0.5$ as a tensor parameter, the equality, $a_+ = a_-$ corresponds to the N-EL and A-EL states assigned above.  However, it is possible to obtain two different energy levels in both the $a_+ >0$ and $a_- >0$ ranges. In the mathematical lines, $a_+ =0.6$ and $a_- =0.4$ values occur for $U_0=0.4$ and these conditions are plotted in Figure  \ref{fig:enter-label3} for the $(n\kappa)=(00)$ ground state, the light-mass in core and heavy-mass in shell.  The optical transitions between the energy levels with identical spins exhibit the blue-shift as the energy differences decrease with increasing well width.

It is arguable how these particle states would fill space or occupy energy levels. In a way, it is possible to fill the space with particle state representations or N-EL with different probabilities of occurrence when two solutions are available. In the other way, principal quantum numbers can be assigned, i.e. the lower and upper energy levels can be given values $n=0$ and $n=1$ respectively. Even if such situations are mathematically possible, they may need to be verified experimentally.
    
    In the above analyses, the shift of the excited energy levels of Dirac particles in the spherical core/shell structure to lower energy levels with reference to the ground state and the increase in energy levels with increasing well width were assigned as ‘anomalous levels’ and thus antiparticle states were detected. As a result of the Coulomb tensor interaction, the existence of two energy levels exhibiting particle behaviour at the same mass is still a puzzle phenomenon in terms of questioning how to occupy the energy levels. 
\bibliography{ref}


\end{document}